\documentclass[11pt]{article}
\usepackage{coling2020}
\usepackage{times}
\usepackage{url}
\usepackage{latexsym}
\usepackage{xcolor}
\usepackage{graphicx}
\usepackage{tipa}
\usepackage{tabularx}
\usepackage{multirow}
\graphicspath{ {./images/} }

\title{A Benchmark for Lease Contract Review}

\author{Spyretta Leivaditi \\
  University of Amsterdam \\
  {\tt spyretta.leiv@gmail.com} \\\And
  Julien Rossi \\
  University of Amsterdam\\
  {\tt j.rossi@uva.nl } \\\And
  Evangelos Kanoulas \\
  University of Amsterdam \\
  {\tt e.kanoulas@uva.nl} \\}

\date{}

\usepackage{hyperref}
\hypersetup{
    colorlinks=true,
    linkcolor=blue,
    filecolor=blue,      
    urlcolor=blue,
    citecolor=black
}

\begin{document}
\maketitle
\begin{abstract}
  
Extracting entities and other useful information from legal contracts is an important task whose automation can help legal professionals perform contract reviews more efficiently and reduce relevant risks. In this paper, we tackle the problem of detecting two different types of elements that play an important role in a contract review, namely entities and red flags. The latter are terms or sentences that indicate that there is some danger or other potentially problematic situation for one or more of the signing parties. We focus on supporting the review of lease agreements, a contract type that has received little attention in the legal information extraction literature, and we define the types of entities and red flags needed for that task. We release a new benchmark dataset of 179 lease agreement documents that we have manually annotated with the entities and red flags they contain, and which can be used to train and test relevant extraction algorithms. Finally, we release a new language model, called ALeaseBERT, pre-trained on this dataset and fine-tuned for the detection of the aforementioned elements, providing a baseline for further research\footnote{Our codebase and model weights are available on \href{https://gitlab.com/spyretta.leiv/lease_contract_review}{Gitlab}}.

\end{abstract}

\section{Introduction}
\label{intro}

A legal contract is a document that describes an agreement between two or more parties that is enforceable by law. A legal contract review, in turn, is a process that includes verifying and clarifying the facts and provisions included in the contract, assessing the contract's feasibility, and predicting potential risks. It is essentially a risk control and management process that aims to prevent damages that may occur during the execution of the contract.

Given this, performing a contract review in a thorough and comprehensive way is essential. At the same time, it is a task that can be time consuming, costly and repetitive; characteristics that make its partial automation highly desirable by both legal professionals and their clients. With that in mind, in this paper we focus on the problem of automating the identification of two different types of contract elements that play an important role in a contract review, namely entities and red flags. 

Entities in a contract can be the contracting parties, dates of different events, amounts of money, specific rights and obligations, and/or related governing laws. Red flags, on the other hand, are terms or sentences that indicate (or even prove)  that there is some danger or other potentially problematic situation for one or more of the signing parties. For example, in a lease agreement the existence of a paragraph or even a sentence that allows the break of the contract before its termination date is automatically a risk for both the lessor and the lessee.

Automatically extracting such elements from contracts is a challenging task for which several systems and methods have been developed in the past years \cite{Chalkidis2017} \cite{DBLP:journals/corr/abs-1805-01217} \cite{5708127}. In this paper, we decided to work on lease agreements, a contract type for which, to the best of our knowledge, no labeled data and/or element extraction system is available. The main novelty of our work is the definition and investigation of the red flagging task in this type of agreement.


In particular, the main contributions of this paper are the following:

\begin{enumerate}

\item The definition of the types of entities and red flags whose automatic detection is necessary for reviewing a lease agreement.

\item The release of a new benchmark dataset of 179 lease agreement documents that we have manually annotated with the entities and red flags they contain, and which can be used to train and test relevant extraction algorithms.

\item The release of a new language model, called ALeaseBERT, that we have pre-trained on lease data and fine-tuned for the aforementioned elements, and which provides a baseline for further research.

\end{enumerate}

The rest of the paper is organized as follows. In the next section we provide a short overview of related approaches to legal information extraction. In section \ref{contract_review} we define the types of contract elements our work supports, while section \ref{dataset} describes the characteristics and annotation methodology of the lease agreement dataset we provide. Section \ref{ALeaseBERT}, in turn, describes the development of AleaseBERT and the results it achieves in identifying the entities and red flags of lease agreements. Finally, in section \ref{conclusions} we summarize the key aspects of our work and we discuss the potential directions it could take in the future.

\section{Related Work}
\label{related}

While to the best of our knowledge there are no datasets or systems for extracting review-related information from lease agreements, there are related approaches that focus on different types of legal documents and different target information.

\newcite{Chalkidis2017}, for example, focus on contract documents and the extraction of elements like the contract's title, its start and termination dates, the contracting parties, the contract's value and others. The dataset they use consists of 3,500 labeled contracts from the U.S. Securities and Exchange Commission \cite{SEC}, while their method combines machine learning (with hand-crafted features and embeddings) and manually written post-processing rules.

Contracts are also analyzed in \cite{5708127}, where a linguistic pattern-based approach is used to identify exceptions in 2,500 business contracts describing services. \newcite{Elwany2019}, in turn, trains and fine-tunes a BERT model on a proprietary corpus of several hundred thousands of legal agreements and uses it to improve the effectiveness of a classifier that detects two types of agreement terms, namely terms that expire after a fixed amount of time and terms that are automatically renewed. 

Another contract-related system is Claudette \cite{DBLP:journals/corr/abs-1805-01217}, which automatically detects and classifies potentially unfair clauses in online terms of service. For that it applies an ensemble of different machine learning algorithms (SVM, HMM, LSTM and others) on a set of 12,000 sentences from 50 on-line consumer contracts. 

Other relevant approaches focus on legislation documents. ~\newcite{chalkidis2019extreme}, for example, consider the problem of extreme multi-label text classification and experiment with several neural classifiers on a set of 57,000 legislation documents from the European Union's public document database (EUR-LEX) that have been annotated with 7,000 concepts from EUROVOC, a multilingual thesaurus maintained by the Publications Office of the European Union. \newcite{10.1145/1165485.1165506}, on the other hand, extracts provisions from legislative texts, along with their type (e.g., repeal, delegation, prohibition, etc.), as well as legal entities involved in these provisions. \newcite{Chalkidis2018} use approx. 120,000 legislation documents from all around the world to train two individual word2vec models for 100-dimensional and 200-dimensional embeddings.

A third group of relevant works extract information from court cases. In \cite{10.1145/3372124.3372128}, for example, the authors analyze the citations found in australian court cases and explore the effectiveness of neural network architectures in determining the category of these citations (neutral, positive, cautionary, negative) as well as their importance. The dataset they use contains a total of 125,000 citations and associated labels, while the method they apply combines BERT and SVM. \newcite{10.5555/2167945.2167948}, in turn, identify and resolve entities referring to judges, attorneys, companies, jurisdictions, and courts in documents describing depositions, pleadings and other trial documents. This is done by training an SVM classifier on based on a manually annotated data of 400 documents.

Also, \newcite{Xiao2018} consider the problem of predicting the judgment results of criminal cases. For that, they implement a number of baseline methods, using a set of 2.6 million criminal cases, published by the Supreme People’s Court of China and annotated with detailed judgement results (including law articles, charges, and prison terms). Finally, prediction of court rulings is also the subject of \cite{sulea2017predicting}, this time on a set of 127,000 cases from the Supreme Court of France. In addition to predicting the ruling, the authors build classifiers to also predict the law area and the period of the cases, using a linear SVM classifier trained on lexical features.

Our work is inspired by the above approaches in the sense that we also build an annotated dataset and train a language model to enable the automatic extraction of information from legal documents. However, it differs significantly in the targeted type of documents (lease agreements) and information to be extracted or identified (lease entities and red flags).


\section{Contract Review: Entities and Red Flags}
\label{contract_review}

\subsection{Entity Extraction}

In a lease agreement the parties agree on the rights and the obligation of a lessor and a lessee.  While there are many types of entities we could possibly extract from such an agreement, our focus in this paper is on fourteen entity types that were indicated by legal professionals specialised in real estate law.
These include  entities about the parties, the property, the terms and rent conditions and important dates and periods.

\begin{itemize}  
    \item \textbf{Entities related to the parties:}
    \textbf{\textit{Lessor information,}} full name, address and phone number of the lessor;
    \textbf{\textit{lessee information,}} full name, address and phone number of the lessee.
    
    \item \textbf{Entities related to the property:}
    \textbf{\textit{Leased space,}} the size of the leased property;
    \textbf{\textit{designated use,}} the predefined use of the property such as "office", "shop", "house", etc.
    
    \item \textbf{Entities related to terms and rent conditions:}
    \textbf{\textit{General terms,}} the Articles and Laws the agreement follows;
    \textbf{\textit{terms of payment,}} the terms and conditions of the money transaction between two parties, such as the amount of rent or deposit;
    \textbf{\textit{VAT obligation,}} the tenant's obligation to pay Value Added Tax;
    \textbf{\textit{indexation rent,}} the amount or percentage the rent increases in specific time periods.
    
    \item \textbf{Entities related to important dates and periods:}
    \textbf{\textit{Start date,}} the date the lease agreement becomes effective, also called effective date;
    \textbf{\textit{signing date,}} the date the agreement is signed, usually different from the effective date;
    \textbf{\textit{expiration lease date,}} the date a lease agreement is terminated;
    \textbf{\textit{notice period,}} the word span that defines the notification period of the cancellation of an agreement from each party;
    \textbf{\textit{rent review date,}} the exact date the rent should be reviewed by the two parties;
    \textbf{\textit{extension period,}} the period both parties agree that the Lease agreement shall be extended for an additional period of time.
    
\end{itemize}

\subsection {Red Flag Detection}

This task attempts to automatically detect parts of the contract that contain one or more risks (called red flags), as well as identify the type of these risks. The list of red flags may vary depending on the scope (financial, regulatory etc) of the review. 
After consultation with legal professionals in real estate law, we selected seventeen red flags, as shown below. The selected risks were those that were most common and first priority for the lawyers and independent to the scope of the contract review. 


\begin{itemize}
    \item \textbf{Red flags related to the contract:}
    \textbf{\textit{Break option,}} a clause that allows the sudden termination of a lease;
    \textbf{\textit{extension period,}} a clause stating the period by which the lease can be extended;
    \textbf{\textit{special stipulations,}} a clause that supplements and, in certain events, modifies or varies, the other provisions of the lease.

    \item \textbf{Red flags related to the landlord's obligations:}
    \textbf{\textit{Compulsory reconstruction,}} a clause about the reconstruction of the leased property in case of complete disaster such as earthquake or fire;
    \textbf{\textit{damage,}} a clause stating how the cost for repairing damages is split among the lessee and the lessor;
    \textbf{\textit{expansion,}} a clause stating the obligation of the Lessor to expand part of the property, at their own cost and expense;
    \textbf{\textit{landlord repairs,}} a clause stating the obligation of a landlord to pay all the expenses of maintenance repairs;
    \textbf{\textit{service charges,}} a clause stating the service costs the lessor charges the lessee to recover their costs in providing services to the building;
    \textbf{\textit{warranties of the owner,}} Written guarantees of the landlord, promising to execute specific property-related actions (e.g. repairs) in a specified period of time;
    \textbf{\textit{guarantee transferable,}} a clause stating that in the event of a sale or other transfer of the building or transfer of the lease, the lessor shall transfer the cash deposit or letter of credit to the transferee, and the lessor shall thereupon be released by the tenant from all liability for the return of such security.

    \item \textbf{Red flags related to the tenant's obligations and permissions:}
    \textbf{\textit{Assignment permitted,}} a clause that stated that the tenant shall have the right to assign the lease contract to its successors or assigns;
    \textbf{\textit{sublease permitted,}} similar to above tenant is allowed to sublease the leased property;
    \textbf{\textit{no obligation to operate,}} a clause that states that the lessee is not obliged to operate the leased property in accordance with the primary intended use;
    \textbf{\textit{reinstatement,}} a clause that states the tenant's obligations with respect to the property's reinstatement in the state it was prior to the beginning of the lease;
    \textbf{\textit{right of first refusal to purchase (ROFR to purchase),}} a clause that gives a tenant the right to buy a leased property before the landlord negotiates any other offers;
    \textbf{\textit{right of first refusal to lease (ROFR to lease),}} a clause that gives a tenant the right to lease a leased property before the landlord negotiates any other offers.
    
    \item \textbf{Miscellaneous red flags:}
    \textbf{\textit{Change of control,}} a clause that provides that where the lessee is a corporate entity and the control of that corporate entity changes as a result of transfer or sale of shares in the corporate entity, then the corporate entity must obtain the lessor’s consent to such change of control.

\end{itemize}

There were also another five red flag types that were taken into account during data collection but we could not provide annotated data for them due to the lack of examples. Those were: \textbf{\textit{termination under a year,}} a clause stating that lease can be terminated less than 12 months from its effective date; \textbf{\textit{rent review,}} a clause used to provide the landlord with an opportunity to review the level of rent payable by a tenant during the term of a lease; \textbf{\textit{indexation,}} a clause stating the annual adjustment of the rent to the cost of living; \textbf{\textit{bank guarantee,}} an agreement between the bank and the lessor that in case the lessee does not pay, the bank will pay the lessor’s claim; \textbf{\textit{lease with CV’s,}} the tenant is a dutch limited partnership company.

Finally, there were two red flag types, namely \textbf{\textit{special stipulation}} and \textbf{\textit{service charges}}, that were not considered during data collection. However, we were advised by a legal professional to include these types in the annotation.

\section{Annotated Dataset}
\label{dataset}

In this section we provide a detailed description of the dataset we designed and produced for the tasks of legal entity extraction and red flag detection. 

\subsection{The Origins of the New Dataset}

The documents we used to create the dataset are publicly available from the U.S. Securities and Exchange Commission \cite{SEC}. The SEC is responsible for overseeing the U.S. securities markets and protecting investors. It provides access to registration statements, periodic financial reports, and other security forms through its electronic data-gathering, analysis, and retrieval database, known as EDGAR. We used EDGAR to retrieve lease agreements since the red flag task focuses on potential risks these agreements may contain. 

\subsection{Choosing Which Contracts To Annotate}


From a pool of 1,631 lease agreements that we retrieved through EDGAR, we chose those that could provide examples of both tasks of this study. Since all lease agreements contain most of the general legal entities, we chose agreements that could provide examples of red flags. We used the BM25 ranking function to estimate the relevance of documents to specific red flag types.

For this process we needed a list of keywords/queries that could indicate the presence of a red flag in the documents. To create this list we asked for the help of two master students in Law and their supervisor. Table 1 shows a sample of this list while the full list can be found in our github repository.

\begin{table}[ht]
\centering 
\begin{tabular}{ |l|l| } 
 \hline
\textbf{Red flag type} & \textbf{Sample keywords} \\
 \hline
 Sublease & sublease, charter, hire, rent out etc. \\ 
 \hline
 ROFR to purchase & right of first refusal to purchase, ROFR to purchase etc. \\ 
 \hline
 ROFR to lease & right of first refusal to lease, ROFR to lease etc. \\ 
 \hline
 As is reinstatement & as is reinstatement, as it is, restore etc. \\ 
 \hline
 Option to purchase & option to purchase, purchase option, right to choose etc. \\ 
 \hline
 No obligations to operate & no obligation to operate, no commitment obligations etc. \\
 \hline
 Bank guarantee & bank guarantee etc.\\ 
 \hline
 Rent review & rent review, review of the rent, revision of the rent etc.\\ 
 \hline
 No transferable security & non transferable security, security is not transferable etc. \\ 
 \hline
 Warranties & Warranties, warranties of the owner, warranties of householder etc. \\ 
 \hline
 Compulsory reconstruction & compulsory reconstruction, reconstruction, destroy, fire etc.\\ 
 \hline
 C.V. & CV, C.V. \\ 
 \hline
 Change of control & ownership, change in management, change of lessor etc. \\ 
 \hline
 Break option & break option, termination, expire etc. \\ 
 \hline
 Termination & termination, limit of, finality etc.\\ 
 \hline
 Indexation & indexation, index, price increase etc. \\ 
 \hline
 Landlord repairs & Landlord repairs, fix, reconstruct etc. \\ 
 \hline
 Damage & destroying, harm, damage etc. \\ 
 \hline
 Expansion & expansion, restructure, developing, remodel etc. \\ 
 \hline
\end{tabular}
\label{tab:bm25_keywords}
\caption{Keywords for red flag types}
\end{table}

For each keyword/query of each red flag type, we retrieved the first 100 most relevant documents. Because it was possible for a lease agreement to have multiple red flags, and for each red flag type we used more than one keyword/query, BM25 gave us multiple duplicates that we removed to end up with 400 unique agreements. From these, we annotated 179 documents.



\subsection{Annotation Effort}

The annotations of the labeled dataset were provided by one law master student in consultation with her supervisor. All four tasks were completed through the same annotation tool by the same annotator. The tool that was used is called TagTog (https://www.tagtog.net), a text annotation editor with which the user can quickly annotate and normalize entities, spans and relations.

\subsubsection {Working with TagTog}

In order to minimize the complexity of the annotation process we decided to treat all our tasks as an entity extraction task. That meant that every legal entity and red flag was considered a TagTog ``Entity''.  

Since red flag task is an identification task and we needed more than just the span of legal risk, we added labels to the ``redflag'' TagTog entity. The labels were the names of the target red flags. This method minimized the annotator's training time and the complexity of the annotation process.

After defining the annotation method, we created a pool of the unique lease agreements collected. From that pool, the annotator selected lease agreements one at a time and started highlighting the requested information.

\subsubsection {The Annotation Process}

The annotation of each document was performed in two phases. In the first phase, the annotators went through the whole document and highlighted the clause/subclause numbering, clause/subclause titles, the definitions and the annex. In this way they had a very quick scan of the content of the document and were better prepared for the next phase. In the second phase, the annotators went through the document a second time, identifying the parts of the text that denoted a legal entity or a red flag. Moreover, they selected the type of red flag from a predefined drop-down list we had created in the tool.

\subsection{Dataset Statistics}

\begin{figure}[ht]
\centering
\includegraphics[width=0.41\columnwidth]{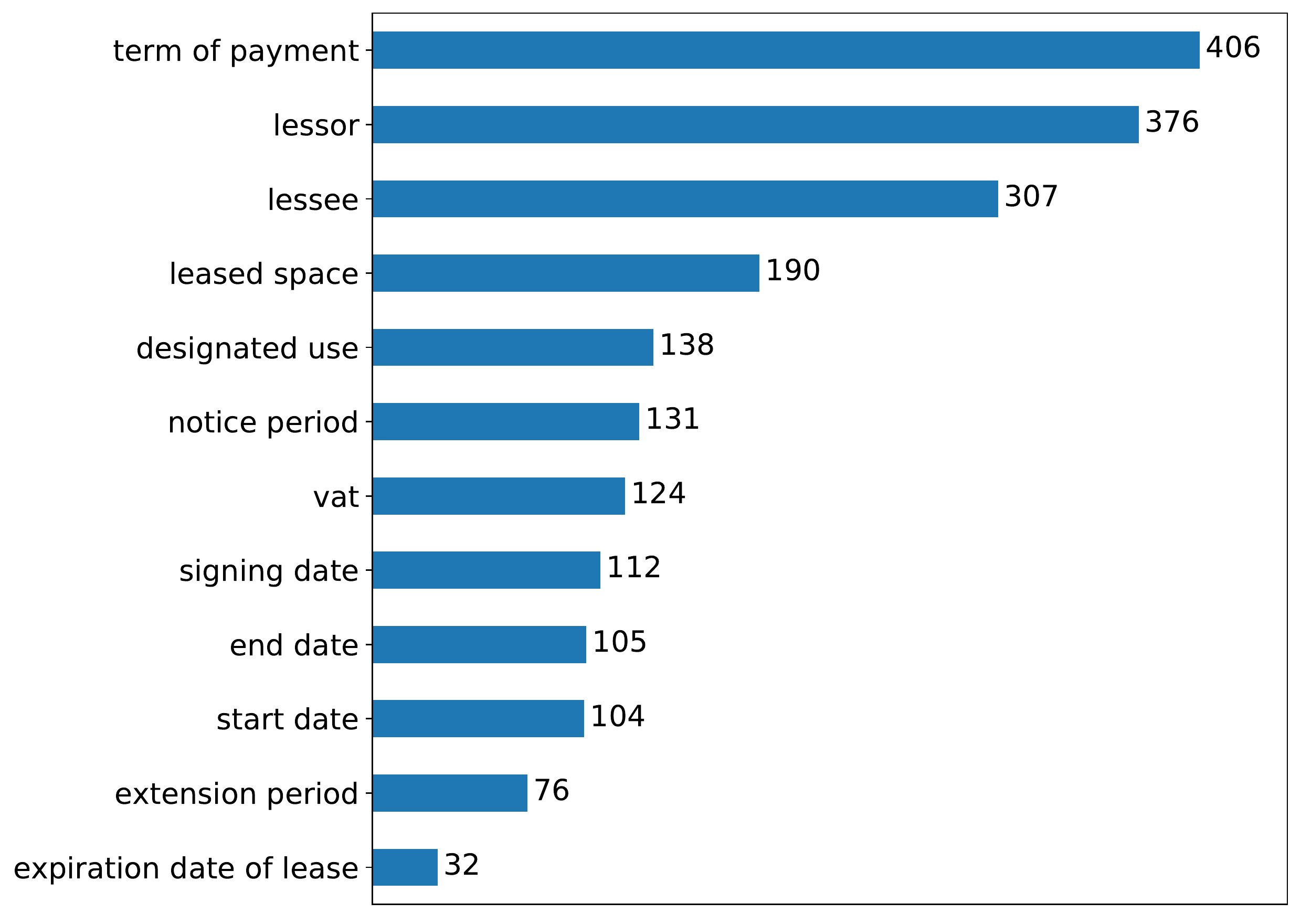}
\includegraphics[width=0.48\columnwidth]{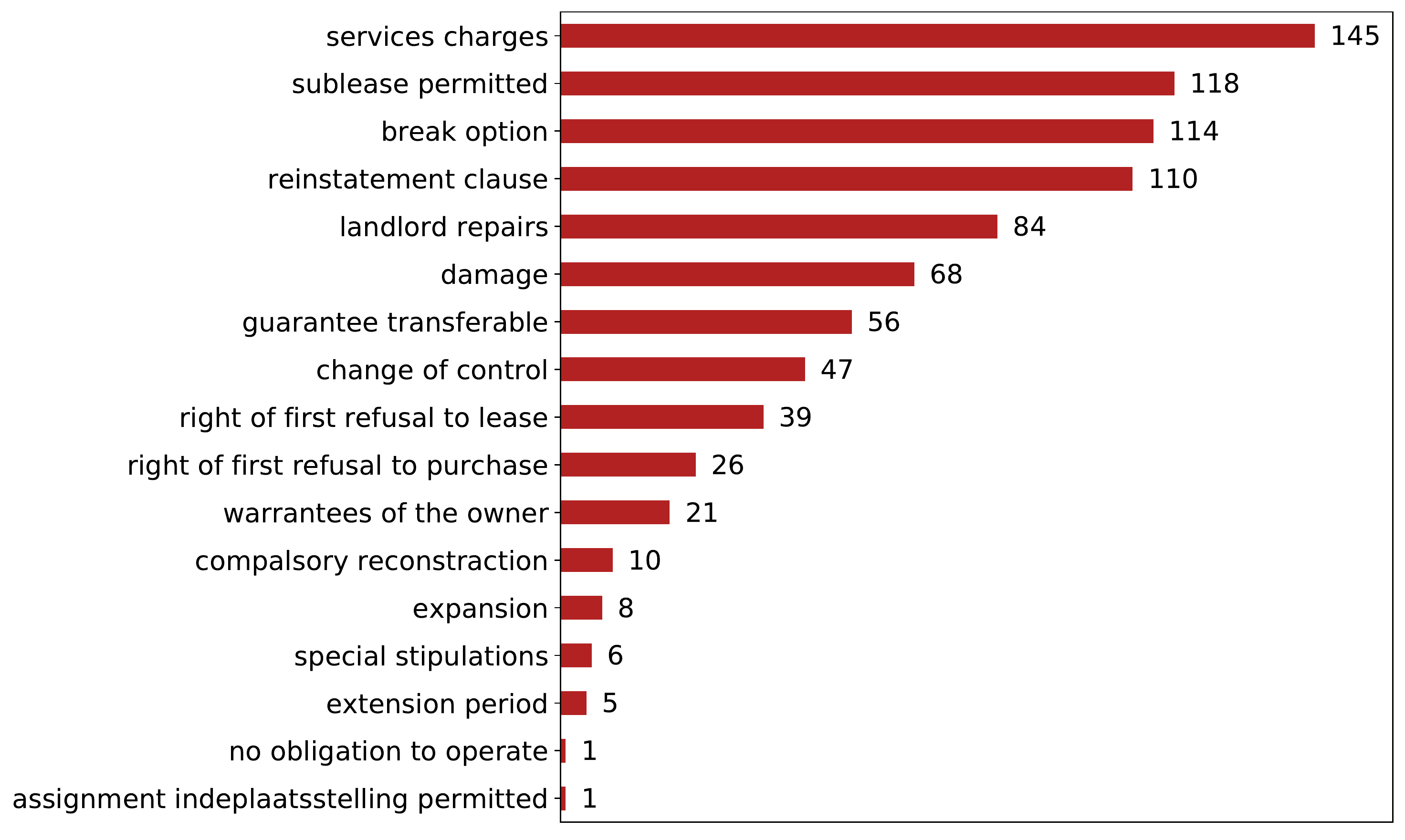}
\caption{Distribution of different types of entities (left) and red flags (right) in the dataset}
\label{fig:entities_redflag_dist}
\end{figure}

The final labeled dataset consists of 179 annotated documents. All documents have red flags, while entities are found in 123 documents. The left chart in Figure \ref{fig:entities_redflag_dist} shows the distribution of the different entity types across all documents, while the right chart does the same for the red flags.

\section{ALeaseBERT [\textschwa \textprimstress lisb\textrevepsilon rt]}
\label{ALeaseBERT}

To enable the automatic detection of entities and red flags, we select generalized language models. These are based on the encoder module of Transformer~\cite{DBLP:journals/corr/VaswaniSPUJGKP17} and have established state-of-the-art performance on many NLP tasks since the first publication of BERT~\cite{DBLP:journals/corr/abs-1810-04805}. 

In particular, the versatility of BERT as a task learner has delivered advances in many NLP tasks in the family of sentence classification or token classification.  \newcite{qiu2020pretrained} surveyed more than 25 different models derived from BERT, published in the 18 months following BERT's publication. For the vast majority of the models, the motivation was either extending BERT's language knowledge to specific professional languages (e.g. BioBert~\cite{10.1093/bioinformatics/btz682}, SciBert~\cite{Beltagy2019SciBERT}) or reducing the computational needs for pre-training and fine-tuning models (e.g. ALBERT~\cite{lan2019albert}, DistilBert~\cite{sanh2019distilbert}). 

In this paper, we contribute an extension of ALBERT, named ALeaseBERT. This is a \texttt{albert-base-v2} model, on which the language modeling task was continued with the lease agreements dataset we described in previous section (1,631 lease contracts, extracted from SEC archive EDGAR), managing to significantly improve the performance of the model in the downstream tasks, as~\cite{rossi2019legal} demonstrated for information retrieval applications in the legal domain.

We make our codebase available through a Github repository as well through huggingface~\cite{Wolf2019HuggingFacesTS}\footnote{After end of anonymity period}


\subsection{Training and Evaluation Methodology}
\label{baselines}

The goal of ALeaseBERT is to enable the extraction of entities and detection of red flags. For both tasks, we split the annotated dataset into training data and validation data, publishing results as observed on the latter. We also established models based on the fine-tuning of ALBERT, in the "ALBERT Base" configuration with embeddings size of 128, all parameters being shared across layers and a total number of trainable parameters of 12 million. We experimented with larger configuration of the same model and concluded that the limited size of our training material did not allow for an efficient learning. 

\subsubsection{Red Flag Detection}
\label{subsub:redflagextraction}

Red flag detection is about identifying contract clauses that should raise alarms, as they represent a risk for a party. In the context of professional search, we consider this task as a total recall task.

We translate this task into a sentence-level ranking task, derived from a binary classifier. Our training material is extracted from the annotations, where contract parts are annotated as being either red flag or neutral. From the original dataset, contracts are split in parts that can be casually identified as sentences. Each sentence is considered as an instance of the positive class if it was identified as, or contained, a red flag. The original annotations classified the red flags in 19 different classes; we will leave the multi-class classification for future work and focus here on the binary classification task.

This is a highly imbalanced dataset, made of 53,232 samples, 51,990 of which belong to the negative class. 
We also provide a strong lexical baseline, based on a grid search within a parameter space that includes feature extraction hyper parameters (e.g. N-grams to consider) as well as machine learning methods for classification commonly associated to text classification: SVM, Logistic Regression and Random Forest. 

For the training, conform to practice, we attach a single dense layer with softmax output as a classification head to the output layer of an ALBERT model, considering the embedding of the \texttt{[CLS]} token as our classifier input. With Adam optimizer~\cite{kingma2014adam}, we train our model for 10 epochs. Observations show that the model has already learned after 3 epochs, while additional training results in over-fitting, which is inline with published literature. 

We evaluate the derived ranker with the mean average precision (MAP), taken as the area under the interpolated precision-recall curve, as well as the interpolated precision for a recall of 0.8.

\subsubsection{Entity Extraction}
\label{subsub:entityextraction}

Entity extraction is an information extraction task, where structured meta-data is extracted from the unstructured text of the corpus, in order to establish a general overview of the corpus. We translate this task into an entity recognition task, and restrict our work to 12 entity types out of the 23 entity types defined in the original annotations. We left apart the entities related to the document structure itself (e.g. clause and subclause numbering, clause and subclause titles) as there is sufficient published work on extracting these entities. We observed inconsistencies in the annotation of the indexation rent or the type of lease; and we also did not consider the entity ``redflag'', as our work has a different approach for its extraction. 

Our training material is extracted from the annotated documents, then translated into CoNLL format, suitable for training.

We consider a named entity recognition model based on ALBERT, while future work might benefit from extending few-shot learning models to this task. We attach a token classification head to the output layer of an ALBERT model, and consider this a token labeling task. With Adam optimizer~\cite{kingma2014adam}, we train our model for five epochs. Inline with previously reported observations, any additional training would result in over-fitting the model to the training data.

We evaluate recall, precision and f1-score for each type of entity, and rely on the weighted average of these measures when excluding the majority class (the default entity type ``O'') to compare models.

\subsection{Results and Discussion}
\label{results}
In this section, we introduce the experimental results we achieved on the tasks described in sections~\ref{subsub:redflagextraction} and~\ref{subsub:entityextraction}, using the models and metrics defined in these sections. We discuss and analyze these results, looking for a validation of our contribution.

\subsubsection{Red Flag Detection}

In this section, we evaluate the models we developed for the task of identifying the red flags contained in contracts from the test dataset.

\begin{table}[ht]
    \begin{tabularx}{\textwidth}{X|lrrr}
    \hline
    Model & MAP & IP@R=0.8\\
    \hline
    Random ranker & 0.0233 & 0.0233 \\
    ALBERT pre-training from scratch & 0.4844 & 0.2622 \\
    TF-IDF 2-grams + Random Forest & 0.4992 & 0.2660 \\
    ALBERT \texttt{albert-base-v2} & 0.5227 & 0.2529 \\
    ALBERT with additional pre-training (our model) & \textbf{0.5733} & \textbf{0.3579} \\
    \hline
    \end{tabularx}
    \caption{Evaluation metrics for the red flag detection task}
    \label{redflag-results}
\end{table}

Table~\ref{redflag-results} shows the results for the task of red flag detection, considered as a ranking task over the sentences of the documents. 
We establish a Random Forest model based on TF-IDF features as the minimum baseline for this task. With regards to the amount of data available for a proper unsupervised language modeling pre-training of an ALBERT model, we conclude that our dataset size falls under the lower boundary, as we observe that a model pre-trained from scratch on our corpus of lease contracts is outperformed by the lexical method. 

We observe that the continuation of pre-training with a domain-specific corpus significantly improves the performance, which we consider in contrast with the results of the model pre-trained on our corpus only. The specifics of lease contract drafting  translate into an update of an already established language model, rather than a complete learning experience in itself. In the light of observations made on the inner workings of BERT~\cite{opensesame}~\cite{bertrediscovers}~\cite{whatdoesbertlookat}, and by extension to ALBERT, we see this phase an adjustment of the ALBERT model with regards to immediate surface features of lease contracts, such as syntax.

In that respect, we observe the limitation of our model to comprehend what constitutes a red flag beyond the language level, as evidenced by the Precision at Recall of 0.8. At 0.35, a human user would have to go through a ranked list of three times the number of red flags in the corpus, in order to retrieve 80\% of those red flags. Based on our findings, red flag detection is a complex task that is far from being sufficiently solved for professional applications.

\subsubsection{Entity Detection}

In this section, we evaluate the models we developed for the task of extracting entities contained in contracts from the test dataset.

\begin{table}[ht]
    \begin{tabularx}{\textwidth}{X|lrrrr}
    \hline
    Model & {} & P & R & F1 & Support \\
    \hline
    \multirow{2}{0.9\columnwidth}{CRF} & macro avg & 0.45 & 0.24 & 0.31 & 11295 \\
                   & weighted avg & 0.53 & 0.37 & 0.43 & 11295 \\
    \hline
    \multirow{2}{0.9\columnwidth}{ALBERT with additional pre-training (our model)} & macro avg & \textbf{0.50} & \textbf{0.35} & \textbf{0.40} & 11295 \\
                                                      & weighted avg & \textbf{0.62} & \textbf{0.48} & \textbf{0.54} & 11295 \\
    \hline
    \end{tabularx}
    \caption{Evaluation metrics for the entity detection task}
    \label{entity-results}
\end{table}

\begin{figure}[t!]
    \centering
    \includegraphics[width=\columnwidth]{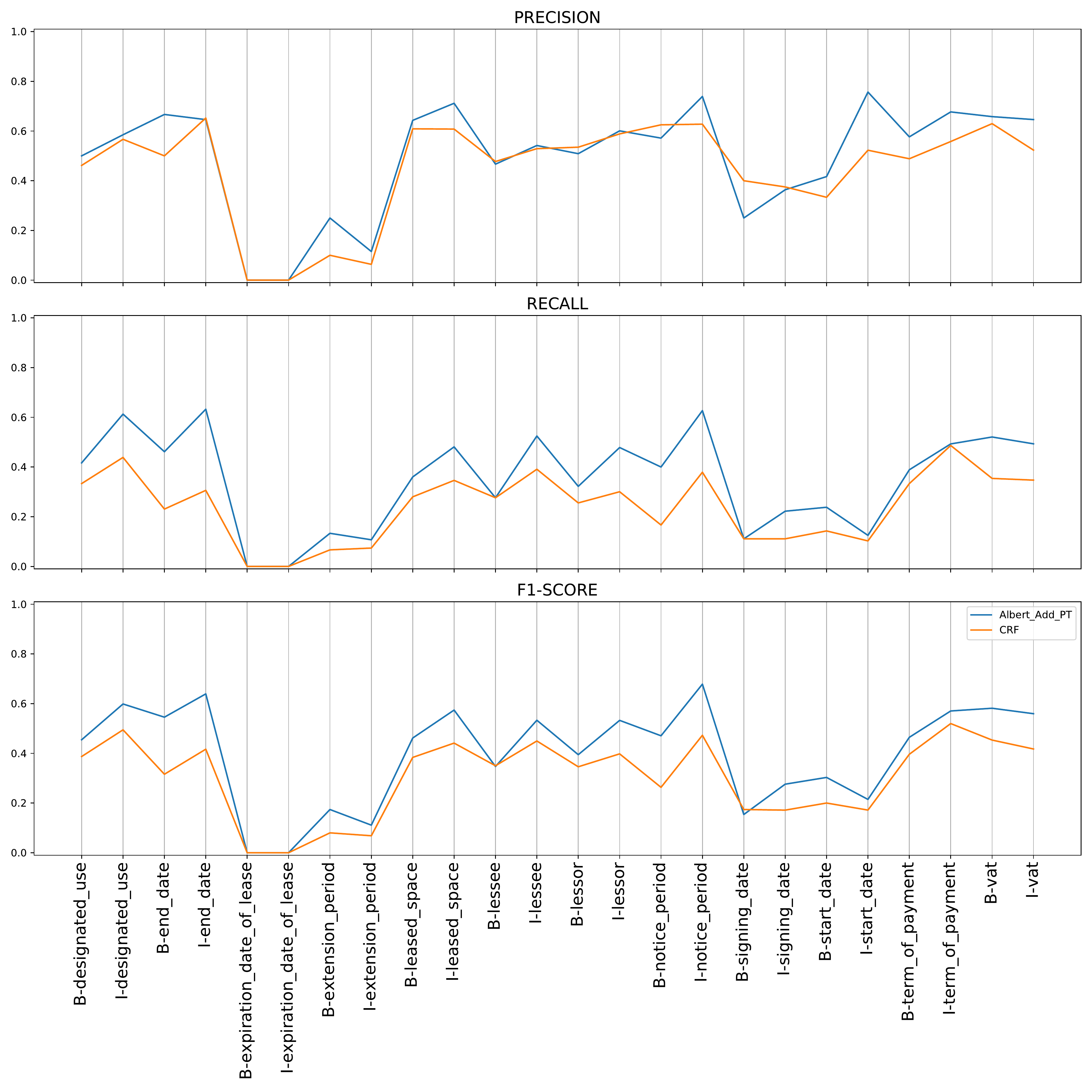}
    \caption{NER by Entity Type}
    \label{fig:entities_ner}
\end{figure}

We report in Table~\ref{entity-results} the main metrics for evaluating our Entity Detection models, and in Figure~\ref{fig:entities_ner} the detailed metrics per entity type. 

We establish a CRF model as the minimum baseline for this task, in accordance with the usage of CRF models as baseline in NER literature, as in~\cite{misawa2017character} or~\cite{kim2019bootstrapping}. While we observe that our model outperforms the baseline, the detailed view of the results show that both models failed to recognize the lease expiration date. We attribute this shortcoming to the limited amount of samples bearing this entity. This creates opportunity for future work on few-shot learning.

Further analysis of the per-entity type performance shows that there are entities with significant support within the dataset for which none of the models could produce a significant performance. In light of our considerations about the red flag detection task, we  see this as a confirmation of the establishment of a complex task by our contribution.

\vspace{60pt}

\section{Conclusions and Future Work}
\label{conclusions}

We introduced the task of detecting red flags and proposed a golden dataset of 179 annotated documents to illustrate this task. Strong baselines for the task of red flag detection were established, based on state-of-the-art generalized language model ALBERT. A language model was pre-trained on lease contracts, and its weights have been published for further usage by the community.

As we experimented with information extraction by performing entity recognition, we identified weaknesses to be investigated for future work: 1) for the red flag detection task, the precision at high recall can be improved for a better professional end user experience; 2) for the entity recognition task, we identified entity types for which we can improve precision and/or recall.

Future work should include mixing signals from entity recognition to improve red flag identification, as well as using state of the art zero-shot of few-shot learning systems for the entity recognition task.

\bibliographystyle{acl}
\bibliography{coling2020}

\begin{thebibliography}{}

\bibitem[\protect\citename{Beltagy \bgroup et al.\egroup
  }2019]{Beltagy2019SciBERT}
Iz~Beltagy, Arman Cohan, and Kyle Lo.
\newblock 2019.
\newblock Scibert: Pretrained contextualized embeddings for scientific text.
\newblock {\em CoRR}, abs/1903.10676.

\bibitem[\protect\citename{Biagioli \bgroup et al.\egroup
  }2005]{10.1145/1165485.1165506}
C.~Biagioli, E.~Francesconi, A.~Passerini, S.~Montemagni, and C.~Soria.
\newblock 2005.
\newblock Automatic semantics extraction in law documents.
\newblock In {\em Proceedings of the 10th International Conference on
  Artificial Intelligence and Law}, ICAIL ’05, page 133–140, New York, NY,
  USA. Association for Computing Machinery.

\bibitem[\protect\citename{Chalkidis and Kampas}2018]{Chalkidis2018}
Ilias Chalkidis and Dimitrios Kampas.
\newblock 2018.
\newblock Deep learning in law: early adaptation and legal word embeddings
  trained on large corpora.
\newblock {\em Artificial Intelligence and Law}, 27:1--28, 12.

\bibitem[\protect\citename{Chalkidis \bgroup et al.\egroup
  }2017]{Chalkidis2017}
Ilias Chalkidis, Ion Androutsopoulos, and Achilleas Michos.
\newblock 2017.
\newblock Extracting contract elements.
\newblock In {\em Proceedings of the 16th edition of the International
  Conference on Articial Intelligence and Law}, pages 19--28.

\bibitem[\protect\citename{Chalkidis \bgroup et al.\egroup
  }2019]{chalkidis2019extreme}
Ilias Chalkidis, Manos Fergadiotis, Prodromos Malakasiotis, Nikolaos Aletras,
  and Ion Androutsopoulos.
\newblock 2019.
\newblock Extreme multi-label legal text classification: A case study in eu
  legislation.
\newblock {\em arXiv preprint arXiv:1905.10892}.

\bibitem[\protect\citename{Clark \bgroup et al.\egroup
  }2019]{whatdoesbertlookat}
Kevin Clark, Urvashi Khandelwal, Omer Levy, and Christopher~D. Manning.
\newblock 2019.
\newblock What does {BERT} look at? an analysis of bert's attention.
\newblock {\em CoRR}, abs/1906.04341.

\bibitem[\protect\citename{Devlin \bgroup et al.\egroup
  }2018]{DBLP:journals/corr/abs-1810-04805}
Jacob Devlin, Ming{-}Wei Chang, Kenton Lee, and Kristina Toutanova.
\newblock 2018.
\newblock {BERT:} pre-training of deep bidirectional transformers for language
  understanding.
\newblock {\em CoRR}, abs/1810.04805.

\bibitem[\protect\citename{Dozier \bgroup et al.\egroup
  }2010]{10.5555/2167945.2167948}
Christopher Dozier, Ravikumar Kondadadi, Marc Light, Arun Vachher, Sriharsha
  Veeramachaneni, and Ramdev Wudali, 2010.
\newblock {\em Named Entity Recognition and Resolution in Legal Text}, page
  27–43.
\newblock Springer-Verlag, Berlin, Heidelberg.

\bibitem[\protect\citename{Elwany \bgroup et al.\egroup }2019]{Elwany2019}
Emad Elwany, Dave Moore, and Gaurav Oberoi.
\newblock 2019.
\newblock Bert goes to law school: Quantifying the competitive advantage of
  access to large legal corpora in contract understanding, 11.

\bibitem[\protect\citename{{Gao} \bgroup et al.\egroup }2012]{5708127}
X.~{Gao}, M.~P. {Singh}, and P.~{Mehra}.
\newblock 2012.
\newblock Mining business contracts for service exceptions.
\newblock {\em IEEE Transactions on Services Computing}, 5(3):333--344.

\bibitem[\protect\citename{Kim \bgroup et al.\egroup
  }2019]{kim2019bootstrapping}
Juae Kim, Youngjoong Ko, and Jungyun Seo.
\newblock 2019.
\newblock A bootstrapping approach with crf and deep learning models for
  improving the biomedical named entity recognition in multi-domains.
\newblock {\em IEEE Access}, 7:70308--70318.

\bibitem[\protect\citename{Kingma and Ba}2014]{kingma2014adam}
Diederik~P. Kingma and Jimmy Ba.
\newblock 2014.
\newblock Adam: A method for stochastic optimization.

\bibitem[\protect\citename{Lan \bgroup et al.\egroup }2019]{lan2019albert}
Zhenzhong Lan, Mingda Chen, Sebastian Goodman, Kevin Gimpel, Piyush Sharma, and
  Radu Soricut.
\newblock 2019.
\newblock Albert: A lite bert for self-supervised learning of language
  representations.

\bibitem[\protect\citename{Lee \bgroup et al.\egroup
  }2019]{10.1093/bioinformatics/btz682}
Jinhyuk Lee, Wonjin Yoon, Sungdong Kim, Donghyeon Kim, Sunkyu Kim, Chan~Ho So,
  and Jaewoo Kang.
\newblock 2019.
\newblock {BioBERT: a pre-trained biomedical language representation model for
  biomedical text mining}.
\newblock {\em Bioinformatics}, 09.

\bibitem[\protect\citename{Lin \bgroup et al.\egroup }2019]{opensesame}
Yongjie Lin, Yi~Chern Tan, and Robert Frank.
\newblock 2019.
\newblock Open sesame: Getting inside bert's linguistic knowledge.
\newblock {\em CoRR}, abs/1906.01698.

\bibitem[\protect\citename{Lippi \bgroup et al.\egroup
  }2018]{DBLP:journals/corr/abs-1805-01217}
Marco Lippi, Przemyslaw Palka, Giuseppe Contissa, Francesca Lagioia,
  Hans{-}Wolfgang Micklitz, Giovanni Sartor, and Paolo Torroni.
\newblock 2018.
\newblock {CLAUDETTE:} an automated detector of potentially unfair clauses in
  online terms of service.
\newblock {\em CoRR}, abs/1805.01217.

\bibitem[\protect\citename{Locke and Zuccon}2019]{10.1145/3372124.3372128}
Daniel Locke and Guido Zuccon.
\newblock 2019.
\newblock Towards automatically classifying case law citation treatment using
  neural networks.
\newblock In {\em Proceedings of the 24th Australasian Document Computing
  Symposium}, ADCS ’19, New York, NY, USA. Association for Computing
  Machinery.

\bibitem[\protect\citename{Misawa \bgroup et al.\egroup
  }2017]{misawa2017character}
Shotaro Misawa, Motoki Taniguchi, Yasuhide Miura, and Tomoko Ohkuma.
\newblock 2017.
\newblock Character-based bidirectional lstm-crf with words and characters for
  japanese named entity recognition.
\newblock In {\em Proceedings of the First Workshop on Subword and Character
  Level Models in NLP}, pages 97--102.

\bibitem[\protect\citename{Qiu \bgroup et al.\egroup }2020]{qiu2020pretrained}
Xipeng Qiu, Tianxiang Sun, Yige Xu, Yunfan Shao, Ning Dai, and Xuanjing Huang.
\newblock 2020.
\newblock Pre-trained models for natural language processing: A survey.

\bibitem[\protect\citename{Rossi and Kanoulas}2019]{rossi2019legal}
Julien Rossi and Evangelos Kanoulas.
\newblock 2019.
\newblock Legal search in case law and statute law.
\newblock In {\em JURIX}, pages 83--92.

\bibitem[\protect\citename{Sanh \bgroup et al.\egroup
  }2019]{sanh2019distilbert}
Victor Sanh, Lysandre Debut, Julien Chaumond, and Thomas Wolf.
\newblock 2019.
\newblock Distilbert, a distilled version of bert: smaller, faster, cheaper and
  lighter.

\bibitem[\protect\citename{SEC}2020]{SEC}
SEC.
\newblock 2020.
\newblock Edgar: Sec database.
\newblock \url{https://www.sec.gov/edgar/search-and-access}.

\bibitem[\protect\citename{Sulea \bgroup et al.\egroup
  }2017]{sulea2017predicting}
Octavia-Maria Sulea, Marcos Zampieri, Mihaela Vela, and Josef Van~Genabith.
\newblock 2017.
\newblock Predicting the law area and decisions of french supreme court cases.
\newblock {\em arXiv preprint arXiv:1708.01681}.

\bibitem[\protect\citename{Tenney \bgroup et al.\egroup }2019]{bertrediscovers}
Ian Tenney, Dipanjan Das, and Ellie Pavlick.
\newblock 2019.
\newblock {BERT} rediscovers the classical {NLP} pipeline.
\newblock {\em CoRR}, abs/1905.05950.

\bibitem[\protect\citename{Vaswani \bgroup et al.\egroup
  }2017]{DBLP:journals/corr/VaswaniSPUJGKP17}
Ashish Vaswani, Noam Shazeer, Niki Parmar, Jakob Uszkoreit, Llion Jones,
  Aidan~N. Gomez, Lukasz Kaiser, and Illia Polosukhin.
\newblock 2017.
\newblock Attention is all you need.
\newblock {\em CoRR}, abs/1706.03762.

\bibitem[\protect\citename{Wolf \bgroup et al.\egroup
  }2019]{Wolf2019HuggingFacesTS}
Thomas Wolf, Lysandre Debut, Victor Sanh, Julien Chaumond, Clement Delangue,
  Anthony Moi, Pierric Cistac, Tim Rault, R'emi Louf, Morgan Funtowicz, and
  Jamie Brew.
\newblock 2019.
\newblock Huggingface's transformers: State-of-the-art natural language
  processing.
\newblock {\em ArXiv}, abs/1910.03771.

\bibitem[\protect\citename{Xiao \bgroup et al.\egroup }2018]{Xiao2018}
Chaojun Xiao, Haoxi Zhong, Zhipeng Guo, Cunchao Tu, Zhiyuan Liu, Maosong Sun,
  Yansong Feng, Xianpei Han, Zhen Hu, Heng Wang, et~al.
\newblock 2018.
\newblock Cail2018: A large-scale legal dataset for judgment prediction.
\newblock {\em arXiv preprint arXiv:1807.02478}.

\end{thebibliography}
\end{document}